# Round Robin: Quantitative Lateral Resolution of PHI XPS microprobes Quantum 2000 / Quantera SXM

U. Scheithauer[a], M. Kolb[b], G. A.M. Kip[c], E. Naburgh[d], J.H.M. Snijders[d]

[a] *82008 Unterhaching, Germany*
  Phone: + 49 8963644143
  E-mail: scht.uhg@googlemail.com

[b] *Airbus Group Innovations, TX2, 81663 Munich, Germany*
  Phone: + 49 8960720460
  E-mail: max.kolb@airbus.com

[c] *Universiteit Twente, MESA+ Nanolab, Postbus 217, 7500AE Enschede, The Netherlands*
  Phone: +31 653845867
  E-mail: G.A.M.Kip@utwente.nl

[d] *Materials Analysis, Philips Innovation Services, High Tech Campus 11, 5656 AE Eindhoven, The Netherlands*
  Phone: +31 402748144
  E-mail: e.p.naburgh@philips.com / j.h.m.snijders@philips.com

## Keywords

*XPS, ESCA, Quantum 2000, Quantera, spatial resolution*

## Abstract

The quantitative lateral resolution is a reliable measure for the quality of an XPS microprobe equipped with a focused X-ray beam. It describes the long tail contributions of the X-ray beam intensity distribution. The knowledge of these long tail contributions is essential when judging on the origin of signals of XPS spectra recorded on small-sized features.

In this round robin test the quantitative lateral resolution of 7 PHI XPS microprobes has been estimated. As expected, the quantitative lateral resolution has significantly larger values than the nominal X-ray beam diameter. The estimated values of the quantitative lateral resolution follow a trend in time: The newer the monochromator of an XPS microprobe so much the better the quantitative lateral resolution.

# 1. Introduction

The outstanding design feature of Physical Electronics (PHI) XPS microprobes Quantum 2000 and Quantera SXM, respectively, is the double focussing ellipsoidally shaped quartz monochromator of the X-ray source. This device monochromatizes the $Al_{k_\alpha}$ radiation and refocuses the X-rays from the Al anode to the sample surface. In this way a variation of the diameter of the X-ray generating electron beam allows to vary the X-ray beam diameter on the sample surface. The PHI XPS microprobes can be operated with nominal X-ray beam sizes between ~ 10µm and 200 µm.

But it has been shown that for a more precise characterization of the X-ray source the long tail contributions of the X-ray beam intensity distribution has to be taken into account [1]. The knowledge of this X-ray beam intensity distribution is essential when judging on XPS measurements recorded on small sample features. Only with this knowledge one can estimate whether low intensity signals are attributed to contaminations on the feature's surface or if they might have originated from the surrounding of the small feature due to such long tail intensity distribution of the X-ray beam.

Following these considerations, the quantitative lateral resolution is defined by the diameter of the circle, for which only 1% of the X-ray beam intensity hits the surface outside this circle, if the X-ray beam is centred in the circle. For the Quantum 2000, instrument no. 78, the quantitative lateral resolution was estimated to be ~ 450 µm [1]. For this XPS microprobe it improves to ~ 190 µm after an upgrade of the monochromator [2]. In this article the results of an inter-laboratory comparison of the quantitative lateral resolution of different PHI XPS microprobes are presented.

# 2. Instrumentation

For the measurements presented here several PHI Quantum 2000 and Quantera SXM X-ray microprobes were used. The details of the X-ray microprobes are listed in tab. 1. The spatial resolution of these XPS microprobes is achieved by the combination of a fine-focused electron beam generating the X-rays on a water cooled Al anode and an ellipsoidal mirror quartz monochromator [3-8], which refocuses the X-rays on the sample surface. This way, the X-ray beam scans across the sample as the electron beam is scanned across the Al anode. Fig. 1 shows a schematic drawing of the principal components of a PHI X-ray microprobe. By controlling the electron beam diameter, nominal X-ray beam diameters between ~ 10 µm up to 200 µm are adjustable. Nominal beam sizes of the Xray beam diameters were estimated using the method defined by the manufacturer: The nominal beam size is the distance between the points at which the signal amplitude is 20% and 80% of the maximum value when the beam is scanned over a material edge [4, 5, 9].

In the XPS instrument, the sample features are identified by X-ray beam induced secondary electron images. These images are generated while the electron beam used for X-ray generation is rastered on the Al anode, which simultaneously scans the X-ray beam across the sample surface.

| instrument | instr. no. / year | type |
|---|---|---|
| 1a | 78 / 2001 | Quantum 2000 |
| 1b | 78 / 2010 | Quantum 2000 / new monochromator |
| 2 | 83 / 2001 | Quantum 2000, upgrade to Quantera |
| 3 | 27 / 1996 | Quantum 2000 |
| 4 | 49 / 1998 | Quantum 2000 |
| 4a | 49 / 2008 | Quantum 2000, upgrade to Quantera / new monochromator |
| 5 | 82 / 2001 | Quantera |

Tab. 1: instruments of round robin

By emittance matching the acceptance area of the energy analyser is synchronized with the X-ray position on the sample. Voltages that are synchronized with the raster of the exciting electron beam are applied to electrostatic deflection plates at the analyzer entrance for this purpose. By a dynamic dispersion compensation, the energy shifts due to energy variations of the X-rays beam are compensated while the beam position is shifted along the disperse direction of the monochromator. The retarding potential of the analyzer input lens is varied according to the disperse direction raster position of the X-ray beam [7].

For sputter cleaning of the samples each instrument is equipped with a differentially pumped $Ar^+$ ion gun. For all samples measured here in a Quantum 2000 / Quantera SXM the incoming X-rays are parallel to the surface normal. In this geometrical situation, the energy analyser take off axis and the ion gun are oriented ~ 45° relative to the sample surface normal.

## 3. Measurement of Quantitative Lateral Resolution

It was shown that "inverse dots", which are currently Pt apertures with different diameters known from electron microscopy, can be utilized to measure the quantitative lateral resolution [1, 9]. Apertures with diameters of 20 … 600 µm are used. Due to its high XPS sensitivity Pt is a good choice regarding measurement time and signal to noise ratio. As shown by the insert of fig. 2, the apertures are mounted on top of a drilling.

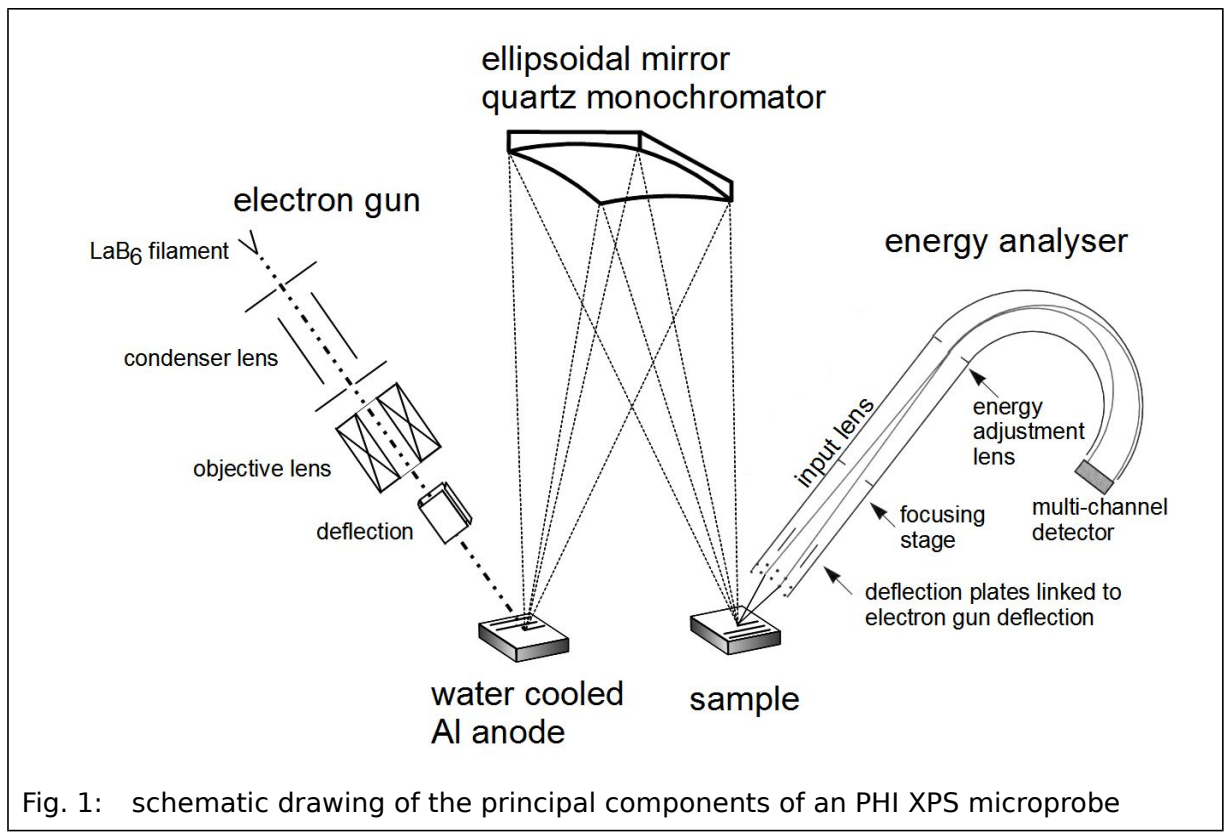

Fig. 1: schematic drawing of the principal components of an PHI XPS microprobe

Due to this geometry, if the apertures are cleaned by $Ar^+$ ion sputtering, redeposited material on the drilling bottom is not visible for the energy analyser because it is geometrically shielded by the aperture itself. For a measurement the focussed X-ray beam is centred in an aperture. In this way only that portion of the X-ray beam intensity distribution contributes to the signal, which comes from outside the radius defined by the aperture diameter. The recorded Pt intensity of the aperture centre measurement is normalised by a second measurement on massive Pt far way from the aperture. This procedure was repeated for all apertures of different diameters using a small or different X-ray beam diameters.

Fig. 2 shows the normalized Pt4f peak areas measured at the aperture centres as function of the aperture diameter for four different-sized X-ray beams. The X-ray beam diameters were estimated utilizing the 20% to 80% criterion when scanning across a knife-edge-type sample. The measured curves point out that two types of X-ray intensities contribute to the signal. In the region, where the apertures are only 2…4 times larger than the beam size, the normalized signal drops down as expected [10]. And additionally a slowly varying background of the X-ray intensity, which decreases with higher aperture diameter, is present. It can be seen, that for apertures having a diameter of 300 µm or more the normalized signal is independent from the beam diameter. The origin of this background is not clarified yet, but most probably it is due to imperfections of the X-ray monochromator (D. Watson, Physical Electronics Inc., Eden Prairie, private communication). If this region is fitted graphically, the quantitative lateral resolution of the instrument 1b (see tab. 1) is estimated to be ~ 190 µm. In other words: If an XPS measurement is done on the centre of a material dot of 190 µm diameter 1 % of the X-ray beam intensity will hit the sample surface outside the dot area. This 1% level is an appropriate parameter to characterize an XPS microprobe. Please realise that the quantitative lateral resolution is independent from the X-ray beam diameter for the 3 smaller-sized X-ray beams used here.

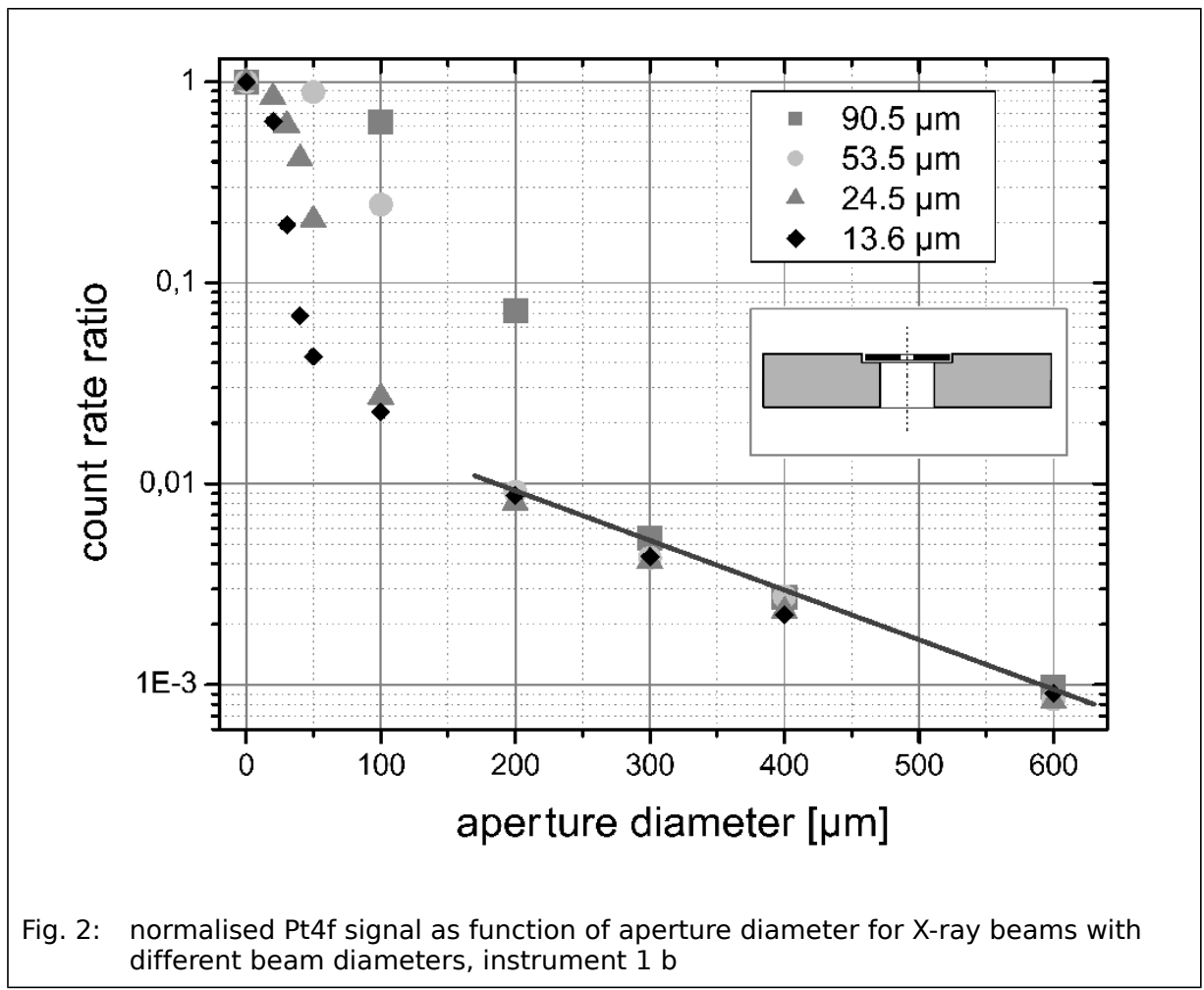

Fig. 2:   normalised Pt4f signal as function of aperture diameter for X-ray beams with different beam diameters, instrument 1 b

The quantitative lateral resolution of this monochromator is ~ 14 times bigger than the diameter of the 13.6 µm X-ray beam. To emphasise, the knowledge about an XPS microprobes lateral resolution function is essential, when judging on measurements recorded on sample features of limited lateral dimensions as bond pads of microelectronic devices, for instance. A bond pad has a lateral size of ~ 70 * 70 µm$^2$. For the monochromator, whose quantitative lateral resolution is shown in fig. 2, ~ 3 % of the X-ray intensity will impinge on the bond pads surroundings and initiate photoelectron emission from this area, if the X-ray beam with a nominal beam size of 13.6 µm is used and centred on the bond pad. These 3% have to be taken into account when interpreting low intensity XPS signals attributed to unexpected elements. Whether these contaminations are on the bond pad surface or are contributions of the bond pad surroundings is indeterminable.

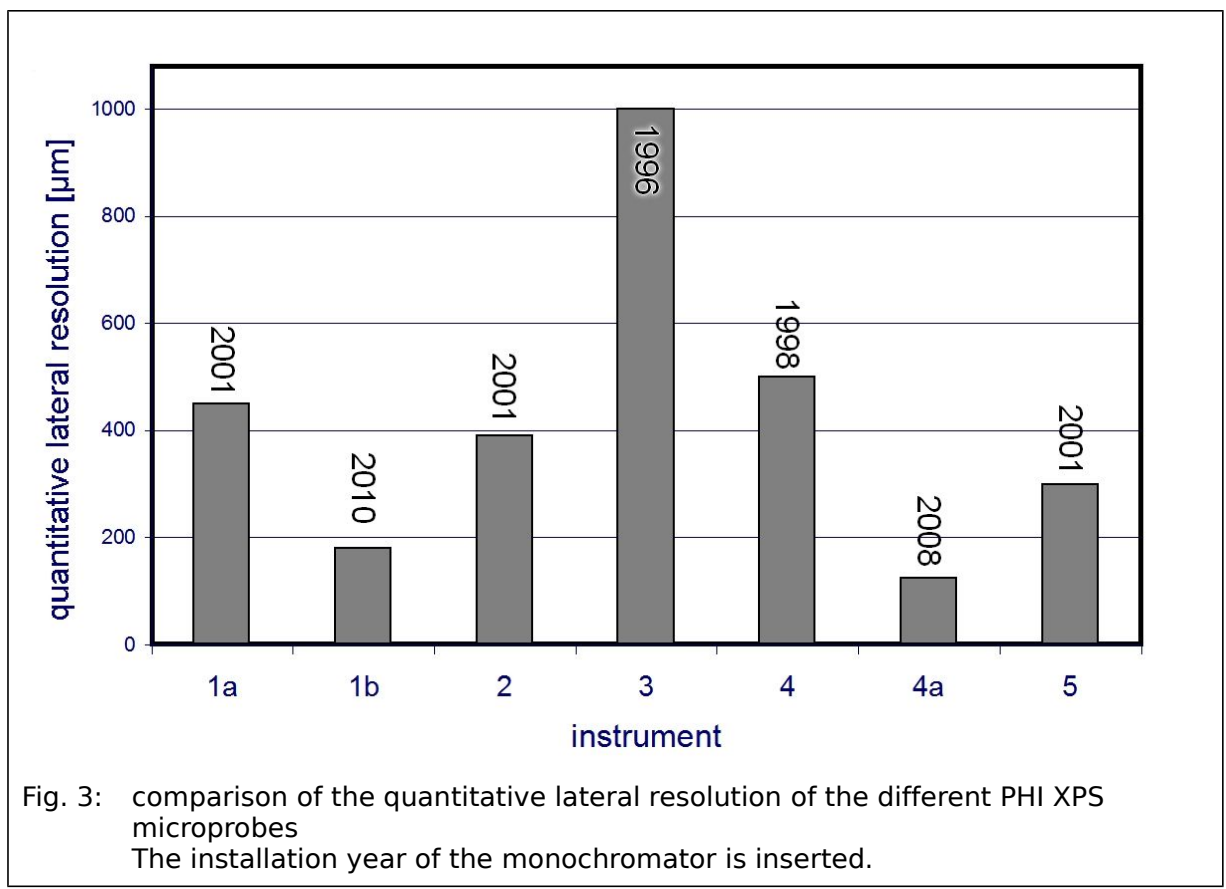

Fig. 3: comparison of the quantitative lateral resolution of the different PHI XPS microprobes
The installation year of the monochromator is inserted.

In this round robin the quantitative lateral resolution of different PHI XPS microprobes has been estimated. The analysed instruments are listed in tab. 1. With all instruments using X-ray beams with small nominal beam sizes in the range of 10 … 20 µm the normalised intensity in the aperture centre of different-sized apertures was measured. Using the 1% level as demonstrated above, the quantitative lateral resolution of the XPS microprobes was estimated. Fig. 3 shows the result for 5 instruments. Two instruments were upgraded with a new monochromator. Therefore these instruments are listed twice. The installation year of the monochromator is inserted. For the instruments, which were characterised in this inter-laboratory comparison, quantitative lateral resolutions between 130 µm and 1000 µm were determined.

## 4. Conclusions

The quantitative lateral resolution describes the long tail intensity distribution of the X-ray beam of an XPS microprobe. Per definition, it is the diameter of the circular area, for which 1% of the X-ray beam intensity hits the sample outside of this area when the X-ray beam is centred in the circular area. Therefore, it is the better characterisation of an XPS instruments lateral resolution performance than the nominal X-ray beam diameter estimated by the 20% to 80% method. As expected, the quantitative lateral resolution has a significantly larger value than the nominal X-ray beam diameter and it is independent from nominal X-ray beam diameters for smaller nominal X-ray beam diameters.

The quantitative lateral resolution of 7 monochromators of PHI XPS microprobes was measured. Values between 130 µm and 1000 µm were determined. The estimated values of the quantitative lateral resolution follow a trend in time: The newer the monochromator so much the better the quantitative lateral resolution. Apparently the instruments manufacturer has improved his skills over the years.

## References

[1] U. Scheithauer, *Quantitative Lateral Resolution of a Quantum 2000 X-ray Microprobe*, Surf. Interface Anal. 40 (2008) 706-709

[2] U. Scheithauer, *Characterisation of the primary X-ray source of an XPS microprobe Quantum 2000*, J. Electron Spectrosc. Relat. Phenom. 193C (2014) 58-62

[3] P.E. Larson, P.W. Palmberg, *Scanning and high resolution XPS and imaging*, Patent-Number: 5315113 (1994)

[4] Physical Electronics Inc., *System Specifications for the PHI Quantum 2000*. Eden Prairie, MN 55344 USA (1999)

[5] Physical Electronics Inc., *PHI Quantera SXM,* Eden Prairie, *MN 55344 USA (2003)*

[6] H. Iwai, R. Oiwa, P.E. Larson, M. Kudo, *Simulation of Energy Distribution for Scanning X-ray Probe*, Surf. Interface Anal. 25 (1997) 202-208

[7] Physical Electronics Inc., *The PHI Quantum 2000: A Scanning ESCA Microprobe.* Eden Prairie, MN 55344 USA (1997)

[8] I.W. Drummond, *Spatial Resolution in X-ray Spectroscopy,* Phil. Trans. R. Soc. Lond. A 354 (1996) 2667-2682

[9] ISO/TR 19319, *Surface chemical analysis - Auger electron spectroscopy and X-ray photoelectron spectroscopy - Determination of lateral resolution, analysis area, and sample area viewed by the analyser,* Geneva, Switzerland *(2013)*

[10] D.R. Baer, M.H. Engelhard, *Approach for determining area selectivity in small-area XPS analysis*, Surf. Interface Anal. 29 (2000) 766-772